\documentclass[twocolumn,pra,aps,showpacs,superscriptaddress,floatfix,showkeys,nofootinbib,10pt]{revtex4-2}

\usepackage[T1]{fontenc}
\usepackage{amsmath,amsfonts,amssymb,amsthm}
\usepackage{dsfont}
\usepackage[toc,page]{appendix}
\usepackage{bbm}
\usepackage{enumitem}
\usepackage{framed}
\usepackage{subfig}
\usepackage{tabularx}
\usepackage{graphicx}
\usepackage{siunitx}
\usepackage{tikz}
\usetikzlibrary{arrows.meta}

\newcounter{rowcntr}[table]
\renewcommand{\therowcntr}{\thetable.(\alph{rowcntr})}

\newcolumntype{N}{>{\refstepcounter{rowcntr}\therowcntr}c}

\AtBeginEnvironment{tabular}{\setcounter{rowcntr}{0}}

\newcommand{\ket}[1]{ | \, #1 \rangle} \newcommand{\bra}[1]{ \langle #1 \, |}

\newcommand{\be}{\begin{equation}} \newcommand{\ee}{\end{equation}}
\newcommand{\ba}{\begin{aligned}} \newcommand{\ea}{\end{aligned}}

\DeclareRobustCommand\openone{\leavevmode\hbox{\small1\normalsize\kern-.33em1}}%

\begin{document}

\title{Mapping correlations between quantum discord and Bell non-locality}

\author{Robert Okula} \email{rbrt.okula@gmail.com}
\affiliation{Department of Algorithms and System Modeling, Faculty of Electronics, Telecommunications and Informatics, Gda\'{n}sk University of Technology, Poland}
\affiliation{Department of Physics, Stockholm University, 106 91 Stockholm, Sweden} 

\author{Adrian Misiak}
\affiliation{Faculty of Electronics, Telecommunications and Informatics, Gda\'{n}sk University of Technology, Poland}

\author{Piotr Mironowicz} 
\affiliation{Department of Algorithms and System Modeling, Faculty of Electronics, Telecommunications and Informatics, Gda\'{n}sk University of Technology, Poland}
\affiliation{International Centre for Theory of Quantum Technologies, University of Gdańsk, Jana Bażyńskiego 1A, 80-309 Gdańsk, Poland}

\date{\today}

\begin{abstract}
We present a numerical framework for the certification and systematic analysis of the relationship between Bell nonlocality and quantum discord. By determining the minimum discord required for a bipartite state to manifest a specific Bell violation, we establish lower bounds for these correlations. We evaluate this methodology across six distinct Bell expressions, comparing their performance through the minimal discord values observed under varying intensities of white noise. Analysis of the resulting optimization landscape suggests the existence of two characteristic classes of optimized states, categorized by their minimized quantum discord.
\end{abstract}


\maketitle

\section{Introduction} \label{sec:introduction}
The discovery of quantum discord revealed that non-locality and entanglement do not exhaust the full diversity of quantum correlations \cite{Ollivier2001}. As a rigorous information-theoretic measure, discord quantifies non-classical bipartite correlations arising from the discrepancy between two classically equivalent forms of mutual information, persisting even in the absence of entanglement \cite{Streltsov2014}.

This metric is remarkably ubiquitous; states with identically zero discord occupy a measure-zero subset within finite-dimensional state spaces, implying that virtually all mixed quantum states natively possess these correlations \cite{ferraro2010almost}. This widespread presence invites further investigation into its operational utility; since discord is nearly universal, its role as a specialized resource requires careful qualification to maintain its discriminatory power in mapping the specific boundaries of quantum advantage within distributed tasks.

Nevertheless, quantum discord possesses concrete operational interpretations. Specifically, discord quantifies the exact markup in entanglement consumption required during quantum state merging when local operations and classical communication are restricted \cite{cavalcanti2011operational}. This identifies discord as a physical resource whose asymmetry directly dictates performance imbalances in dense coding scenarios. Furthermore, recent derivations of analytical lower bounds for geometric quantum discord and extended partial transpose criteria facilitate the practical detection of these correlations in higher-dimensional systems \cite{yiding2024notequantumdiscord}.

The meaning of quantum discord also extends to thermodynamics, where it determines the efficiency discrepancy between quantum and classical Maxwell’s demons during work extraction \cite{zurek2003quantum}, and to information theory, where it drives the non-classical locking of correlations, a phenomenon where minimal classical communication unlocks vast amounts of hidden information, even in the absence of entanglement \cite{divincenzo2004locking}. Recent analytical derivations of lower bounds for geometric quantum discord and extended partial transpose criteria now facilitate the practical detection of these subtle correlations within higher-dimensional systems. These roles are mathematically unified through comprehensive state-space boundary evaluations \cite{modi2012classical}. Physically, in solid-state architectures like coupled double quantum dots, geometric discord demonstrates superior robustness against thermal and Coulomb noise compared to standard entanglement \cite{wang2025quantum}.

Most importantly, however, the usability of quantum discord lies in its ability to capture quantum advantages in information processing tasks where entanglement is either absent or insufficient \cite{Streltsov2014}. It has been identified as the figure of merit for mixed-state protocols such as remote state preparation \cite{Dakic2012}. In this task, separable (non-entangled) states possessing non-zero discord can surprisingly outperform certain entangled states, with the process's efficiency being bounded from below by the geometric measure of discord \cite{Dakic2012}. Similarly, in entanglement distribution, the amount of entanglement that can be successfully transferred between parties is strictly limited by the relative entropy of discord of the carrier particle \cite{Streltsov2012}. In the transmission of correlations, discord represents the unavoidable loss of information when a quantum system is transmitted via classical measurement outcomes; perfect classical transmission is only possible if the initial state has zero discord \cite{Streltsov2013}. It also serves as a method for considering quantum correlations for mixed states \cite{brodutch2017}. It has been also shown, using quantum discord, that separable states can still possess a quantum advantage, manifested in protocols such as deterministic quantum computation with one qubit~\cite{Datta2008}.

To investigate the correspondence between entanglement and quantum discord, this paper introduces a numerical framework for certifying and systematically analyzing the relationship between Bell non-locality and discord. Specifically, we investigate the lower bounds of discord by determining the minimum value required for a bipartite state to manifest a given Bell violation. Utilizing this approach, we characterize six distinct Bell inequalities, comparing their performance through the minimal discord values observed under varying intensities of white noise. The results presented herein offer new insights into the interplay between different forms of quantum correlations.

The paper is structured as follows: Section \ref{sec:discordmethod} details the numerical methodology, including the parametrization of bipartite states, the selection of Bell inequalities, and the computational protocol for discord calculation. Section \ref{sec:discordresults} then presents the findings of our analysis, including the aforementioned certification of the minimal discord values, as well as the discussion on a potential method for classifying states based on their discord.

\subsection{Discord measure definition}
Let us consider a bipartite quantum system $\rho^{AB}$ composed of subsystems $\rho^{A}$ and $\rho^B$. The aggregate correlations within this system are quantified by the quantum mutual information, defined as:
\begin{equation}
    I(\rho^{AB}) = S(\rho^A) + S(\rho^B) - S(\rho^{AB}).
\end{equation}
In the context of quantum discord, the von Neumann entropy $S$ is traditionally restricted to two-level systems ($d=2$). Accordingly, this paper adopts a qubit-based framework for all subsequent discord evaluations. While mutual information encompasses the totality of correlations, an alternative perspective arises from generalizing the classical notion of information gain, which is the reduction in uncertainty regarding subsystem $A$ following a measurement on $B$. This quantity is expressed as:
\begin{equation}
    J(\rho^{AB})_{\{\Pi^{1,y}_i\}} = S(\rho^A) - S(A|\{\Pi^{1,y}_i\}),
\end{equation}
where 
\begin{equation} \label{eq:neumanncond}
    S(A|\{\Pi^{1,y}_i\}) = \sum_i p_i S(\rho_i^{A})
\end{equation}
represents the conditional entropy of $A$ given the outcomes of a measurement performed on $B$. 

While $I$ and $J$ are equivalent for classical systems, they diverge for the quantum states. This discrepancy stems from the invasive nature of quantum measurements, which irreversibly perturb the state and collapse the quantum coherence between subsystems. Because local measurements cannot capture all quantum correlations, $J$ is typically strictly less than $I$. Quantum discord is therefore defined as the minimal discrepancy between these two quantities, optimized over all possible measurement bases:
\begin{equation} \label{eq:discord}
    D(\rho^{AB}) = I(\rho^{AB}) - \max_{\{\Pi^{0,y}_i\}} J(\rho^{AB}),
\end{equation}
where $\{\Pi^{0,y}_i\}$ is the set of projective measurements.

\subsection{Bell non-locality as a measure}
The Bell's theorem posits a scenario involving two spatially separated observers, Alice and Bob, and a referee who distributes an entangled bipartite quantum state (e.g., polarization-entangled photons). Upon receipt, the parties perform local measurements by randomly selecting from sets of observables $\{M^{0,x}\}$ and $\{M^{1,y}\}$. For binary observables, $M^{p,x} = \Pi_+^{p,x} - \Pi_-^{p,x} = 2\Pi_+^{p,x} - \mathbb{I}$. The CHSH operator~\cite{PhysRevLett.23.880} is defined as a linear combination of the measurement expectations:
\begin{equation} \label{eq:chsh}
    \mathfrak{B} = \langle C(0,0) \rangle + \langle C(0,1) \rangle + \langle C(1,0) \rangle - \langle C(1,1) \rangle.
\end{equation}
where the correlation operator $C(x,y)$ is given by:
\begin{equation} \label{eq:correlator}
\begin{split}
    C(x,y) &= M^{0,x} \otimes M^{1,y} \\
    &= 4 \Pi_+^{0, x} \Pi_+^{1, y} - 2 \Pi_+^{0, x} - 2 \Pi_+^{1, y} + \mathbb{I}.
\end{split}
\end{equation}
Under the assumptions of \textit{local realism}, where properties exist independently of observation and measurement choices lack mutual influence, the expression \eqref{eq:chsh} is constrained by a classical bound (which we denote as $|\mathfrak{B}|_{\mathcal{L}}$) of $|\mathfrak{B}_{CHSH}|_{\mathcal{L}} = 2$.

Conversely, entangled quantum systems violate these assumptions; local measurements irreversibly alter the global state. While individual correlators $|\langle C(x,y) \rangle|$ can reach unity, the structure of quantum operators prevents the simultaneous maximization of all four terms in the CHSH expression. Consequently, the individual modules are limited to $1/\sqrt{2}$, yielding \textit{Tsirelson's bound}~\cite{cirelson_quantum_1980} (which we denote as $|\mathfrak{B}|_{\mathcal{Q}}$) of $|\mathfrak{B}_{CHSH}|_{\mathcal{Q}} = 2\sqrt{2}$ for maximally entangled states.

Any violation of the classical bound indicates the presence of entanglement, with the magnitude of violation quantifying its degree. Beyond CHSH, diverse Bell expressions have been developed~\cite{PhysRevLett.61.662, PhysRevA.87.020302, PhysRevLett.129.150403, PhysRevA.78.032112} to detect partial separability~\cite{Gisin_1998, Seevinck_2002, PhysRevLett.88.230406} or to optimize noise resistance using Frank-Wolfe algorithms~\cite{Designolle_2023, PhysRevA.96.012113, brierley2017convexseparationconvexoptimization}.

\section{Characterization for entangled states}  \label{sec:discordmethod}
We analyze a bipartite scenario where Alice and Bob share a generalized two-qubit mixed state. Utilizing a purification-based parametrization, we represent the mixed state as a reduced density matrix of a higher-dimensional pure state, thereby spanning the complete space of two-qubit states \cite{kong2022mixedstateparametrizationtwoqubit}.

To establish the density matrix, first, the basis has to be set. We use a basis constructed from single-qubit unitary transformations, which are parameterized by local rotation angles, as follows:
\begin{align}
|d_0\rangle &\equiv |\phi\phi'\rangle, & |d_1\rangle &\equiv |\phi_\perp\phi_\perp'\rangle, \nonumber \\
|d_2\rangle &\equiv |\phi\phi_\perp'\rangle, & |d_3\rangle &\equiv |\phi_\perp\phi'\rangle;
\label{eq:basis}
\end{align}
where 
\begin{equation}
\begin{split}
    \ket{\phi} &= \cos\left(\frac{\theta}{2}\right) e^{-i\frac{\psi}{2}} \ket{0} + \sin\left(\frac{\theta}{2}\right) e^{i\frac{\psi}{2}} \ket{1}, \\
    \ket{\phi_\perp} &= -\sin\left(\frac{\theta}{2}\right) e^{-i\frac{\psi}{2}} \ket{0} + \cos\left(\frac{\theta}{2}\right) e^{i\frac{\psi}{2}} \ket{1}, \\
    \ket{\phi'} &= \cos\left(\frac{\theta'}{2}\right) e^{-i\frac{\psi'}{2}} \ket{0} + \sin\left(\frac{\theta'}{2}\right) e^{i\frac{\psi'}{2}} \ket{1},\\
    \ket{\phi'_\perp} &= -\sin\left(\frac{\theta'}{2}\right) e^{-i\frac{\psi'}{2}} \ket{0} + \cos\left(\frac{\theta'}{2}\right) e^{i\frac{\psi'}{2}} \ket{1};
\end{split}
\end{equation}
which are parametrized by the angles $\theta, \psi, \theta', \psi'$.

The density matrix is formed by an orthonormal set of entangled basis vectors $\{ \ket{|e_k} \}$. The primary vector, $\ket{e_0}$, is parameterized by the entanglement angle $\chi$ and phase $\zeta$ ($\chi, \zeta \in [0, 2\pi)$). The remaining vectors $|e_k\rangle$ are explicitly derived by mixing the orthogonal complement $|\Psi_\perp\rangle = -e^{-i\zeta}q_-|0\rangle + q_+|1\rangle$ with the basis states \eqref{eq:basis}:
\begin{subequations}
\label{eq:explicit_param}
\begin{align}
\ket{e_0} &= q_+ \ket{d_0} + e^{i\zeta} q_- \ket{d_1}, \\
\ket{e_1} &= c_{21}\ket{\Psi_\perp} + s_{21}(c_{32}\ket{d_2} + s_{32}\ket{d_3}), \\
\ket{e_2} &= -c_0 \bar{s}_{21}\ket{\Psi_\perp} + (c_0 \bar{c}_{21}c_{32} - s_0 \bar{s}_{32})\ket{d_2} \nonumber \\
            &\quad + (c_0 \bar{c}_{21}s_{32} + s_0 \bar{c}_{32})\ket{d_3}, \\
\ket{e_3} &= \bar{s}_0 \bar{s}_{21}\ket{\Psi_\perp} - (\bar{s}_0 \bar{c}_{21}c_{32} + \bar{c}_0 \bar{s}_{32})\ket{d_2} \nonumber \\
            &\quad - (\bar{s}_0 \bar{c}_{21}s_{32} - \bar{c}_0 \bar{c}_{32})\ket{d_3}.
\end{align}
\end{subequations}
The parameters $c_{ij}, s_{ij}$ equal:
\begin{equation}
c_{ij} = \cos\left(\frac{\theta_{ij}}{2}\right) e^{-\frac{i\psi_{ij}}{2}}, \quad
s_{ij} = \sin\left(\frac{\theta_{ij}}{2}\right) e^{\frac{i\psi_{ij}}{2}},
\label{eq:mixing_params}
\end{equation}
where $\theta_{ij} \in [0, 2\pi)$ and $\psi_{ij} \in [0, 2\pi)$ are the rotation angles parameterizing the unitary transformations. The entanglement coefficients for the Schmidt decomposition are given by:
\begin{equation}
q_+ = \cos\chi, \quad q_- = \sin\chi,
\label{eq:entanglement_params}
\end{equation}
with $\chi$ determining the initial entanglement of the pure state core $|e_0\rangle$.

Finally, the mixed state $\rho$ is assembled as the weighted sum of the projectors of these entangled basis vectors:
\begin{equation}\label{eq:rho_sum}
\rho = \sum_{k=0}^{3} \mu_k |e_k\rangle\langle e_k|,
\end{equation}
where the coefficients $\mu_k$ correspond to the eigenvalues of the density matrix and satisfy the normalization condition $\sum \mu_k = 1$, which allows us to parametrize only 3 out of four $\mu_k$ values, while the fourth one is a value derived from \eqref{eq:rho_sum}. Consequently, $\mu_k \in [0,1]$.

The full quantum state is therefore parametrized using 15 separate optimization variables: $\mu_0, \mu_1, \mu_2, \theta, \psi, \theta', \psi', \theta_0, \psi_0, \theta_{21}, \psi_{21}, \theta_{32}, \psi_{32}, \chi, \zeta$.

An input vector $x \in \mathbb{R}^{N_D}, \{ x_i \}_{i=0}^{N_D-1}$ is defined such that the first fifteen components, $\{ x_i \}_{i=0}^{14}$, uniquely determine the density matrix $\rho$ according to the described purification scheme.

\subsection{Parametrization of the Bell operator}
\label{ss:parametrizationofbellinequalities}
To characterize the non-local properties of the generated states, we perform a non-linear optimization over the joint space of state configurations and local measurement settings, constrained by the violation of various Bell inequalities. The remaining components of $\{ x_i \}^{N_D-1}_{i=15}$ characterize the local measurement settings for Alice and Bob. We consider generalized bipartite scenarios where Alice chooses from $N_A$ observables $\{M^{0,k}\}_{k=0}^{N_A-1}$ and Bob from $N_B$ observables $\{M^{1,l}\}_{l=0}^{N_B-1}$. To eliminate redundant degrees of freedom associated with rotational invariance, we fix Alice's first measurement setting to the Pauli operator $M^{0,0} \equiv \sigma_z$.
The remaining observables are fully variational. Each setting $O \in \{M^{0,k>0}, M^{1,l}\}$ is an observable parameterized by polar and azimuthal angles $(\theta, \phi); \theta, \phi \in [0,2\pi)$ on the Bloch sphere:
\begin{equation}
O(\theta, \phi) = \begin{pmatrix}
\cos \theta & e^{-i\phi}\sin \theta \\
e^{i\phi}\sin \theta & -\cos \theta
\end{pmatrix}.
\end{equation}

\subsection{Parametrization of the discord calculation}
\label{ss:parametrizationofdiscordcalculation}
To evaluate \eqref{eq:discord}, we must maximize the classical correlations over all possible local measurements on subsystem $B$. Since the local entropy $S(\rho^A)$ is invariant under operations on $B$, this optimization is equivalent to finding the measurement basis that minimizes the conditional entropy $S(A|\{\Pi^{1,y}_i\})$.

In the bipartite case, we optimize over the set of measurements $\{\Pi^{1,y}_i\}$, which are parameterized by the Bloch sphere angles $\theta_d \in [0, \pi]$ and $\phi_d \in [0, 2\pi)$. The orthogonal basis states defining these measurements are:
\begin{equation}
    \ket{u} = \begin{pmatrix} \cos\frac{\theta_d}{2} \\ e^{i\phi_d}\sin\frac{\theta_d}{2} \end{pmatrix}, \quad
    \ket{v} = \begin{pmatrix} \sin\frac{\theta_d}{2} \\ -e^{i\phi_d}\cos\frac{\theta_d}{2} \end{pmatrix}.
\end{equation}
The corresponding projectors acting on subsystem $B$ are defined as $\Pi^{1,(\theta_d, \phi_d)}_0 = \ket{u}\bra{u}$ and $\Pi^{1,(\theta_d, \phi_d)}_1 = \ket{v}\bra{v}$. Following the calculation of conditional entropy via Eq.~\eqref{eq:neumanncond}, we implement a numerical strategy to find the minimum value. Specifically, the parameters $\theta_d$ and $\phi_d$ are integrated into the global non-linear optimization problem detailed in Section~\ref{ss:parametrizationofbellinequalities}. This increases the total dimension of the search space, which in the end equals $N_D = 15 + 2(N_A-1) + 2N_B + 2$.

\section{Results}  \label{sec:discordresults}
To characterize the boundary between non-classical correlations and Bell non-locality, we perform a numerical minimization of the quantum discord for two-qubit systems under a strict constraint of a Bell value $|\mathfrak{B}|_p = p |\mathfrak{B}|_{\mathcal{Q}}$, where $p \in (0,1]$ parametrizes the constraint by the fraction of the maximal quantum Bell value, Tsirelson's bound. For example, for CHSH and for $p = \frac{1}{\sqrt{2}}$, the value comes down to the local bound $|\mathfrak{B}_{CHSH}|_{p=1/\sqrt{2}} = \frac{1}{\sqrt{2}} |\mathfrak{B}_{CHSH}|_{\mathcal{Q}} = \frac{1}{\sqrt{2}} \cdot 2 \sqrt{2} = 2 = |\mathfrak{B}_{CHSH}|_{\mathcal{L}}$. One of the constraints is thus that the Bell value for the optimized state and measurements $\tilde{|\mathfrak{B}|}$ needs to stay very close to the Bell value for a given $p$:
\begin{equation}
    |\mathfrak{B}|_p - \epsilon \leqslant \tilde{|\mathfrak{B}|} \leqslant |\mathfrak{B}|_p + \epsilon,
\end{equation}
where $\epsilon$ is a small tolerance for a deviation. For our considerations, deviation tolerance equals $\epsilon = 10^{-3}$.

To ensure that the minimization algorithm respects the restrictions mentioned, especially the normalization condition $\sum \mu_k = 1$, we employ the COBYQA \cite{cobyqa} algorithm. The most important characteristic of this method is that it sticks strictly to user-specified bounds, making it ideal for optimizing functions that remain undefined outside those bounds.

For complex regimes where local minima may emerge, we utilize a basin-hopping \cite{Wales_1997} global optimization routine. This, however, could lead to a problem with maintaining strict bounds. The basin-hopping method performs a perturbation of the optimized vector, without paying attention to the bounds defined in the minimization. If these bounds and constraints are exceeded, the minimization will be started in the range of states that are not valid quantum states (especially if the normalization condition $\sum \mu_k = 1$ is violated), leading to the possibility of negative discord. To counter that, we have introduced two separate solutions: first, a reimplementation of the step function has been provided, which does not allow exceeding the defined bounds and the normalization condition. Additionally, we have introduced a step rejection function that, after the minimization for a single basin-hopping step, verifies the correctness of the quantum state, rejecting the result in case of exceeding the bounds.

\subsection{Discord for canonical Bell expressions}
\label{ss:discordforestablishedbellinequalities}
In this section, we present the results of the non-linear Discord function optimization for the three groups of Bell expressions. First – Clauser-Horne-Shimony-Holt \cite{PhysRevLett.23.880} expression and modified CHSH \cite{Mironowicz_2013}:
\begin{align}
\label{eq:modchsh}
    \mathfrak{B}_{Mod} &= \langle C(1,2) \rangle + \langle C(1,3) \rangle + \langle C(2,1) \rangle \\
    &+ \langle C(2,2) \rangle - \langle C(2,3) \rangle.
\end{align}
The Tsirelson's bound \cite{cirelson_quantum_1980} for CHSH equals $|\mathfrak{B}_{CHSH}|_{\mathcal{Q}} = 2 \sqrt{2}$ and for \eqref{eq:modchsh} it equals $|\mathfrak{B}_{Mod}|_{\mathcal{Q}} = 1 + 2 \sqrt{2}$. We also utilize Braunstein-Caves $BC_3$ and $BC_5$ expressions, which were introduced in \cite{PhysRevLett.61.662} and they represent a family of expressions parametrized by several binary measurement settings for two parties:
\begin{equation}
\label{eq:bcn}
\begin{split}
    \mathfrak{B}_{BC_n} &= \langle C(1,1) \rangle + \langle C(1,2) \rangle + \langle C(2,2) \rangle + \ldots\\
    &+ \langle C(n-1,n-1) \rangle + \langle C(n-1,n) \rangle \\
    & + \langle C(n,n) \rangle - \langle C(n,1) \rangle,
\end{split}
\end{equation}
where $n \geqslant 2$ and the Tsirelson's bound equals $|\mathfrak{B}_{BC_n}|_{\mathcal{Q}} = 2 n \cos \left( \frac{\pi}{2n} \right)$. Finally, we utilize $I_1/I_2$ expressions, defined in \cite{Mironowicz_2013}, in a form of:
\begin{equation}
\label{eq:I1}
\begin{split}
    \mathfrak{B}_{I_1} &= \langle C(1,2) \rangle - \langle C(1,3) \rangle -\langle C(2,1) \rangle - \langle C(2,2) \rangle \\
    & + \langle C(3,1) \rangle + \langle C(3,3) \rangle + \langle C(4,1) \rangle,
\end{split}
\end{equation}
and
\begin{equation}
\label{eq:I2}
\begin{split}
    \mathfrak{B}_{I_2} &= - \langle C(1,2) \rangle + \langle C(1,3) \rangle + \langle C(2,1) \rangle \\
    &+ \langle C(2,2) \rangle + \langle C(2,3) \rangle + \langle C(3,2) \rangle \\
    &- \langle C(3,3) \rangle + \langle C(4,1) \rangle + \langle C(4,2) \rangle \\
    &+ \langle C(4,3) \rangle.
\end{split}
\end{equation}
Their Tsirelson's bounds are $|\mathfrak{B}_{I_1}|_{\mathcal{Q}} = 1 + 6 \cos(\frac{\pi}{2})$ and $|\mathfrak{B}_{I_2}|_{\mathcal{Q}} = 2 + 4 \sqrt{2}$ respectively.

\begin{figure}[htbp]
    \centering
	\includegraphics[width=\linewidth]{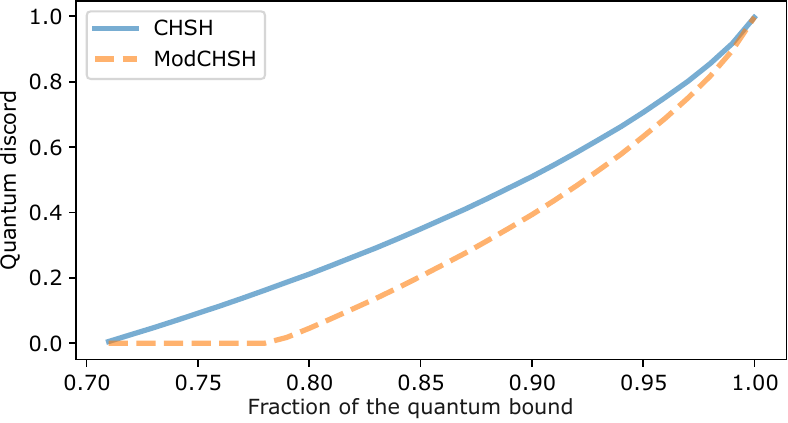}
	\caption{Lower bounds on quantum Discord for CHSH inequality, as well as \eqref{eq:modchsh}, as a function of the fraction $p$ of the quantum bounds $|\mathfrak{B}_{CHSH}|_{\mathcal{Q}}$ or $|\mathfrak{B}_{Mod}|_{\mathcal{Q}}$, respectively.}
	\label{fig:chshdiscord}
\end{figure}

For CHSH and \eqref{eq:modchsh}, the results of the optimization are presented in Figure \ref{fig:chshdiscord}. For the modified Bell expression \eqref{eq:modchsh}, it is visible that all quantum correlations, including those measured by quantum discord, vanish at the classical limit $|\mathfrak{B}|_{\mathcal{L}}$. This phenomenon is also present for CHSH and for other Bell expressions presented. As for the relation between CHSH and \eqref{eq:modchsh}, it can be noticed that the shape of the discord function is the same for both; however, due to \eqref{eq:modchsh} reaching $|\mathfrak{B}|_{\mathcal{L}}$ for higher fractions $p$, the function is compressed along the X axis.

\begin{figure}[htbp]
    \centering
	\includegraphics[width=0.9\linewidth]{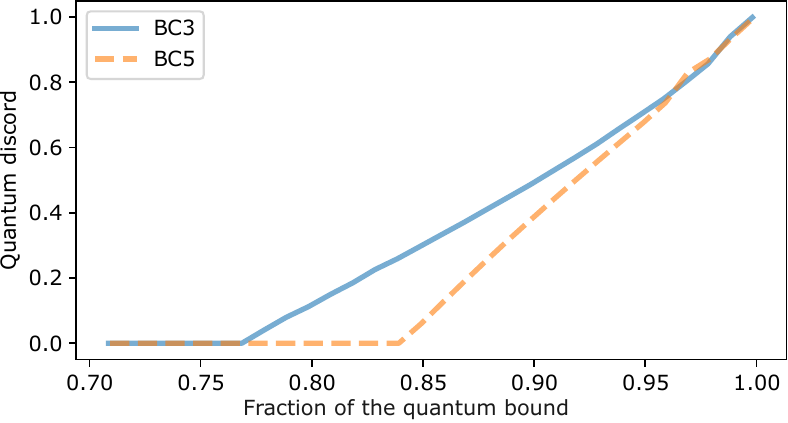}
	\caption{Lower bounds on quantum Discord for the family of Braunstein-Caves inequalities, as a function of the fraction $p$ of the quantum bounds $|\mathfrak{B}_{BC_n}|_{\mathcal{Q}}$.} 
	\label{fig:bcndiscord}
\end{figure}

For the Braunstein-Caves family of chained Bell expressions, the results of the optimization are presented in Figure \ref{fig:bcndiscord}. For the higher fractions $p$, it can be noticed that for both expressions, Discord stays very close; any differences are likely to result solely from the imperfections of non-linear optimization. For $p < 0.95$ however, the graphs clearly deviate, with the $BC_5$ achieving lower discord values and losing any correlation for the fraction of the quantum bound close to $p = 0.85$, which corresponds to the local bound $|\mathfrak{B}_{BC_5}|_{\mathcal{L}}$. This phenomenon is almost identical to the one presented by CHSH and \eqref{eq:modchsh}.

\begin{figure}[htbp]
    \centering
	\includegraphics[width=\linewidth]{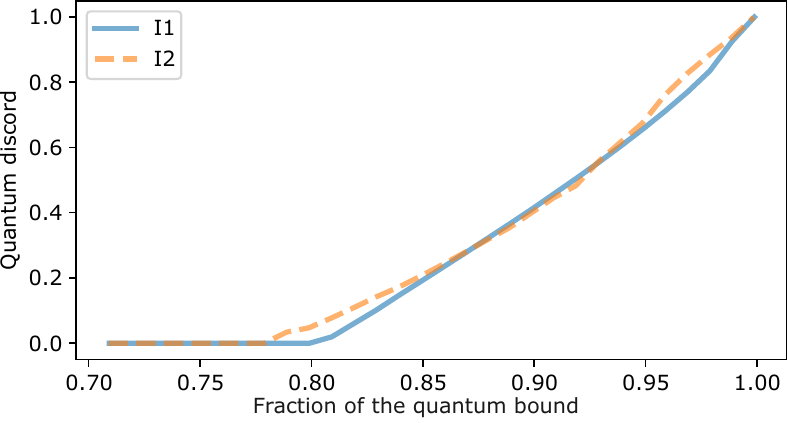}
	\caption{Lower bounds on quantum Discord for the $I_1$ and $I_2$ inequalities, a function of the fraction $p$ of the quantum bounds $|\mathfrak{B}_{I_1}|_{\mathcal{Q}}$ or $|\mathfrak{B}_{I_2}|_{\mathcal{Q}}$, respectively.} 
	\label{fig:indiscord}
\end{figure}

For the $I_n$ family of chained Bell expressions, the results of the optimization are presented in Figure \ref{fig:indiscord}. It can be noticed that for these expressions, the minimized discord plots for both $I_1$ and $I_2$ are very close, and they almost overlap to some extent. For certain noise fractions $p$, $I_2$ exceeds $I_1$. At high fractions ($p>0.95$), this behavior mirrors the phenomenon observed in the Braunstein-Caves family, where the discord values are so nearly identical that discrepancies likely arise from non-linear optimization limits. However, for $p<0.88$, the higher discord values for $I_2$ appear to be an intrinsic feature of the expression itself. The $I_2$ expression achieves its local bound $|\mathfrak{B}|_\mathcal{L}$ for lower fraction, around $p \approx 0.78$. Thus, it shows some level of quantum correlation (in this case, entanglement) also for noise values for which $I_1$ is entirely classical.

\begin{figure}[htbp]
    \centering
	\includegraphics[width=\linewidth]{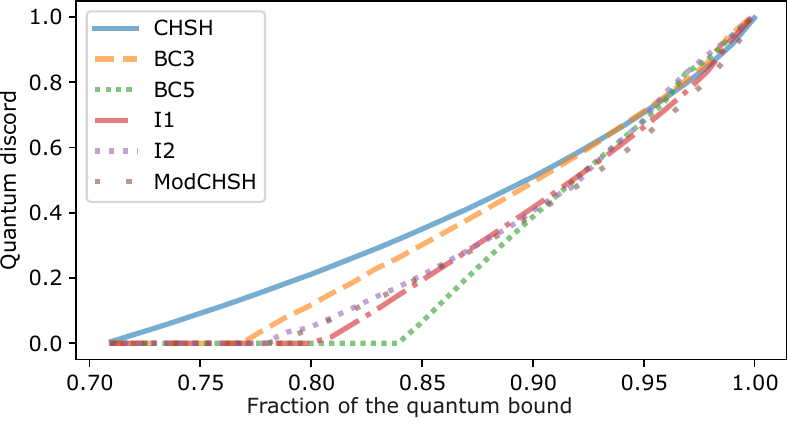}
	\caption{The summary of discord optimization results for all analyzed Bell expressions, as a function of the fraction $p$ of the quantum bound $|\mathfrak{B}|_{\mathcal{Q}}$.} 
	\label{fig:discordcomparison}
\end{figure}

In Figure~\ref{fig:discordcomparison}, we present the comparison of discord values optimized for all six Bell expressions analyzed. The highest minimal discord value can be observed for CHSH for the whole spectrum of noise values, whereas the lowest minimal discord is represented by $BC_5$. In general, for a larger number of measurement settings available for the parties, our discord lower bound decreases. Within the groups of expressions that represent the same number of settings (e.g., $I_1$ and $I_2$), the differences are minimal: they represent roughly the same level of quantum correlation. However, it can be noticed that \eqref{eq:modchsh} shows a level of quantum correlations close to the $I_n$ family, while it allows two fewer measurement settings for Alice and the same number for Bob.

\subsection{Structural Patterns in Discord Minimization}
For the purpose of this analysis, around fifty numerical experiments were performed for each of the Bell expressions, with different optimization parameters such as: initial vector (fully randomized, established for a state where its value is close to its quantum bound $|\mathfrak{B}|_{\mathcal{Q}}$ or with an initial vector taken from the state optimized for a previous fraction $p$), number of basin-hopping iterations, step size, and if it uses the default stepping method or our custom made.

The functions presented in Section \ref{ss:discordforestablishedbellinequalities} represent the minimum taken through all of the numerical experiments, separately for every fraction $p$ of the quantum bound $|\mathfrak{B}|_{\mathcal{Q}}$. However, by combining all the results for a given Bell expression onto a single graph, we can observe additional features of the numerical experiments performed, allowing us to indicate certain clues about the optimized states and the optimization process.

\begin{figure}[htbp]
    \centering
	\includegraphics[width=\linewidth]{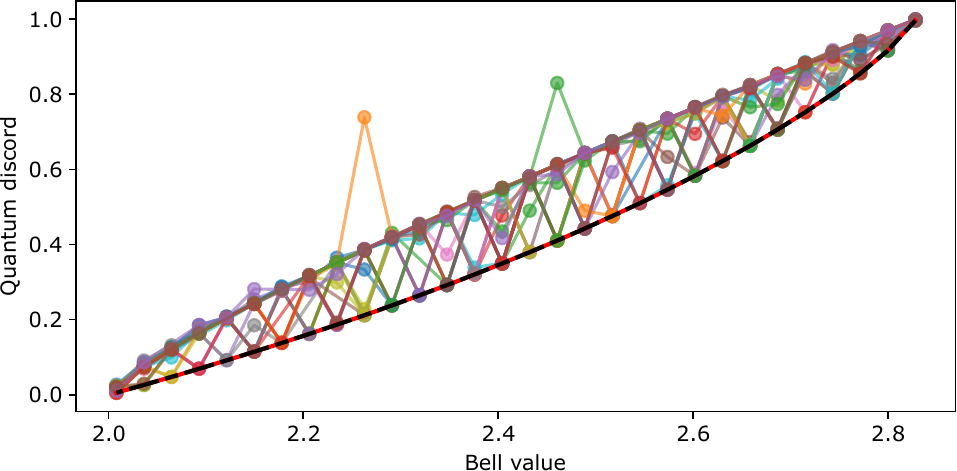}
	\caption{Combination of the results of numerical experiments undertaken to minimize the Discord function for the CHSH expression, as a function of the value of the expression (Bell value).} 
	\label{fig:chshalldiscord}
\end{figure}

In Figure~\ref{fig:chshalldiscord}, we present the aforementioned combined graph for the numerical experiments performed to minimize the Discord function for the CHSH expression. A certain pattern is clearly visible. A distinct pattern emerges, with most of the obtained quantum discord\index{quantum discord} results bounded by two clear curves. One is the minimization curve used in the previous analysis, and the other is its symmetric counterpart (up to optimization error): a concave upper boundary that perfectly inverts the lower convex curve to enclose the data. These two lines indicate the existence of two classes of strong local minima for the optimized states, and thus, they suggest the existence of two distinct classes of quantum states that are more likely to occur.

\begin{figure}[htbp]
    \centering
	\includegraphics[width=\linewidth]{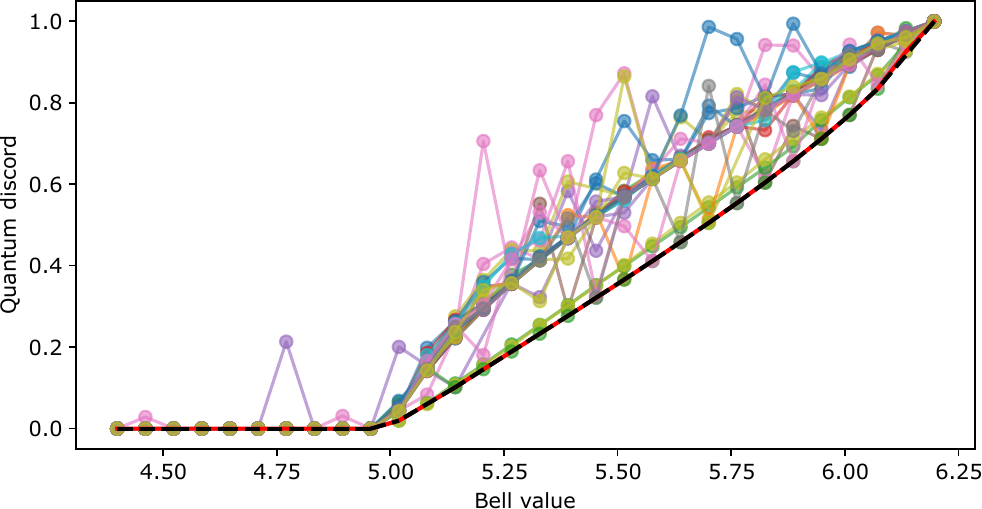}
	\caption{Combination of the results of numerical experiments undertaken to minimize the Discord function for the $I_1$ expression, as a function of the value of the expression (Bell value).} 
	\label{fig:i1alldiscord}
\end{figure}

In Figure~\ref{fig:i1alldiscord}, we combine the plots for the numerical experiments performed to minimize the Discord function for the $I_1$ expression. Here, the two classes of states that we have mentioned before are also visible; however, additional observations can also be made. First, due to the greater number of deviations from the two classes of results, we can observe that the optimization for this Bell expression contains more local minima, which makes minimization more difficult. Furthermore, there is a visible subclass of results that converges to a straight line and has slightly higher discord values than strict minimization. This suggests that the global form of the optimized function must have an additional strong local minimum in the vicinity of the optimal value.

\begin{figure}[htbp]
    \centering
	\includegraphics[width=\linewidth]{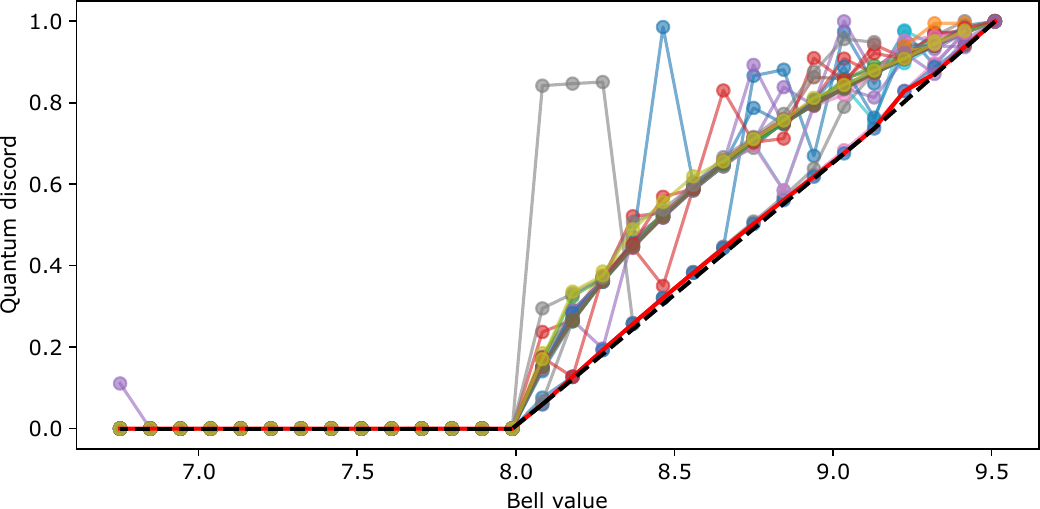}
	\caption{Combination of the results of numerical experiments undertaken to minimize the Discord function for the $BC_5$ expression, as a function of the value of the expression (Bell value).} 
	\label{fig:bc5alldiscord}
\end{figure}

In Figure~\ref{fig:bc5alldiscord}, we combine the plots for the numerical experiments performed to minimize the Discord function for the $BC_5$ expression. Here, the convexity of the minimum function is very close to a straight line, which deviates from the results presented in Figure~\ref{fig:chshalldiscord}, and suggests that all the optimization was stuck in the additional close strong local minimum observed in Figure~\ref{fig:i1alldiscord}. Nevertheless, this result might also be caused by other features resulting from the form of the expression. Without a full characterization of the function, it is not possible to state with certainty that values below the optimized ones are achievable.

\section{Conclusions}  \label{sec:conclusions}
We conducted a comprehensive numerical analysis of quantum discord by performing several dozen non-linear optimization experiments for several Bell expressions to identify the global minimum across a spectrum of noise levels. The optimization of non-linear discord functions across diverse Bell inequalities reveals a consistent physical behavior wherein quantum correlations, even those quantified by discord, vanish entirely as the system reaches its classical limit. Across all analyzed Bell expressions, quantum discord consistently vanishes at their respective classical limits. Furthermore, an increased number of measurement settings generally lowers the minimal discord bound, with CHSH maintaining the highest correlations. Furthermore, optimized quantum discord is visibly bounded by symmetric convex and concave curves, indicating two probable classes of quantum states. Nevertheless, much remains to be discovered regarding the link between quantum discord, Bell violations, and universal discord characterization. This study demonstrates that classifying quantum states by their discord values might uncover compelling new properties within these classes.

\section{Acknowledgements}
We acknowledge the use of a computational server financed by the Foundation for Polish Science (IRAP project, ICTQT, contract no. 2018/MAB/5/AS-1, co-financed by the EU within the Smart Growth Operational Programme). Nonlinear optimization was achieved using SciPy \cite{2020SciPy-NMeth}, using COBYQA as the optimization solver \cite{rago_thesis,cobyqa}.

\bibliographystyle{plainnat}
\bibliography{shannon}

\end{document}